\documentclass[a4paper,fleqn]{cas-dc}

\usepackage[authoryear,longnamesfirst]{natbib}

\usepackage{extarrows} 
\def\tsc#1{\csdef{#1}{\textsc{\lowercase{#1}}\xspace}}
\tsc{WGM}
\tsc{QE}

\begin{document}
\let\WriteBookmarks\relax
\def\floatpagepagefraction{1}
\def\textpagefraction{.001}

\shorttitle{Uniqueness}    

\shortauthors{Rajk\'{o} and Razaq}  

\title [mode = title]{Uniqueness in multivariate curve resolution, re-tuned}



%

\author[1]{R\'{o}bert Rajk\'{o}}[orcid=0000-0002-6234-658X]

\cormark[1]


\ead{rajko.robert@uni-obuda.hu}

\ead[url]{https://rajko.kemometria.hu/}

\credit{Methodology, Software, Validation, Formal analysis, Investigation, Resources, Data curation, Writing (original draft, review and editing), Visualization, Supervision, Project administration}

\affiliation[1]{organization={University Research and Innovation Center (EKIK), \'{O}buda University},
            addressline={B\'{e}csi \'{u}t 96/b}, 
            city={Budapest},
            citysep={}, 
            postcode={H-1034}, 
            country={Hungary}}

\author[1]{Rabeea Razaq}[orcid=0009-0002-1183-1267]


\ead{razaq.rabeea@uni-obuda.hu}


\credit{Methodology, Validation, Formal analysis, Investigation, Resources, Data curation, Writing (review and editing)}


\cortext[1]{Corresponding author}



\begin{abstract}
There are many misunderstandings about the term and interpretation of uniqueness in multivariate curve resolution tasks. In CAC2026 Tarragona, it turned out that even mathematicians do not properly construe Manne's theorems. So we decided to provide this re-tuned summary, using several original quotes, their explanations, and newly created illustrative figures to get the points across.  
This contribution then explicitly offers the following: (i) a unified interpretation of Manne’s conditions, DBU, GRU, minimal constrained duality, and particular-solution-based criteria; (ii) newly constructed visual illustrations, including a counterexample; and (iii) a practical workflow for diagnosing uniqueness in nonnegative bilinear MCR.
\nocite{*} 
\end{abstract}


\begin{highlights}
\item Manne's theorems (are not sufficient, but necessary) - the full story
\item Data-based uniqueness (DBU)
\item General rule for uniqueness (GRU)
\item Minimal constrained duality vs complementarity/coupling theorems 
\item Particular solution based approach in chemometrics and in mathematics
\end{highlights}


\begin{keywords}
Minimal constrained duality; Uniqueness in MCR;  Manne's theorems, DBU, GRU; Particular solution based uniqueness criteria
\end{keywords}

\maketitle

\section{Introduction}\label{Intro}





First of all we have to decipher the terms and definitions used by~\cite{MANNE1995} who introduced the three resolution theorems. In his abstract, he claims the following: {\it ''The theory of model-free resolution of data matrices from hyphenated chromatography is considered. In this experimental technique the elution products of chromatography are simultaneously analyzed by spectroscopy. A simple example shows that
the resolution of spectra for embedded peaks and concentration profiles for embedding peaks is impossible without use of modelling data. General conditions are given for when such resolution can and cannot be performed. These results include and extend those previously obtained by Maeder and Malinowski.''} Thus, the term resolution means here the unique decomposition of the profile. If the profile is not unique, it cannot be resolved. 

This statement is confirmed by the following sentence: {\it ''After a presentation of basic concepts of hyphenated chromatography, a simple two-component example is presented which cannot be uniquely resolved.''} Manne's aim was definitely not to prove just the existence of a feasible solution, but formulating the conditions how to obtain resolvable profiles, i.e., the criteria for getting unique profile solution. The closing sentence of his Introduction section confirms this: {\it ''A final section discusses the use of additional information in order to obtain unique resolution.''} His final section was the third one and the title was {\it ''Resolution theorems''}. 

The three theorems, later shown to be incomplete, follows:
\begin{description}
\item[Theorem 1.] If all interfering compounds that appear inside the concentration window of a given analyte also appear outside this window, it is possible to calculate the concentration profile of the analyte.
\item[Theorem 2.]  If for every interferent the concentration window of the analyte has a subwindow where the interferent is absent, then it is possible to calculate the spectrum of the analyte.
\item[Theorem 3.] For a resolution based only upon rank information in the chromatographic direction the conditions of Theorems 1 and 2 are not only sufficient but also necessary.
\end{description} 

In the original chromatographic context, Manne’s use of ‘resolution’ is naturally interpreted as recovery of the chemically meaningful, uniquely determined profile rather than merely the computation of one feasible factorization, e.g., the texts {\it ''it is possible to calculate the concentration profile of the analyte''} and {\it ''it is possible to calculate the spectrum of the analyte''} mean the unique resolution of the profile strictly, and not just to be able to calculate a feasible solution from possible many. However, subsequent analyses have shown that the stated profile-based conditions, even when interpreted in this stronger sense, are not sufficient in general to guarantee uniqueness. Their applicability therefore requires examination together with the relevant local-rank, selectivity, and data-geometry conditions.

This paper is organized as follows: first data-based uniqueness (DBU) is explained, with which, it can be proved that even if the Manne's theorems are satisfied, in general, the solution may not be unambiguous. Secondly, the general rule for uniqueness (GRU) is described. At third, duality concept and its misunderstanding will be depicted. Last, chemometrically and mathematically, the particular solution based approach is described. 

Thus this contribution is explicitly offering the following: (i) a unified interpretation of Manne’s conditions, DBU, GRU, minimal constrained duality, and particular-solution-based criteria; (ii) newly constructed visual illustrations, including a counterexample; and (iii) a practical workflow for diagnosing uniqueness in nonnegative bilinear MCR.


\section{Data-based uniqueness (DBU)}\label{DBU}

\cite{RAJKO2015} defined data-based uniqueness as a theorem in two equivalent forms in general, i.e., the theorem was formulated in such way that is independent of the component numbers:
\begin{description}
	\item[Form 1:] If the Borgen plot (or its generalization for arbitrary dimensions) contains a point belonging to the coincidence of vertices of the inner and outer polygon (inner and outer polyhedrons), and, as	a result, this point will be the matching vertices of the two Borgen triangles (all Borgen simplices), then the point signifies the unique solution for the component.
	\item[Form 2:] If a component has a selective window in one direction/way/mode and considering the other direction/way/mode, this component has a sub-window in which the contribution of this
	component is zero while all interfering compounds appear inside this window, then the component can be uniquely decomposed in
	this second direction/way/mode.
\end{description} 

\cite{RAJKO2015} stated the following {\it''Differences between the data-based uniqueness and profilebased uniqueness were described, and we have shown that Manne’s theorems are not sufficient in general. It means that the “unique”
solution obtained by using local rank information and Manne’s theorems may be incorrect, and this fact needs further investigations. It seems that the data-based uniqueness theorem is the correct unification and combination of Manne’s Theorem 1 and 2.''}

\begin{figure*}
	\centering
	\includegraphics[width=0.95\textwidth]{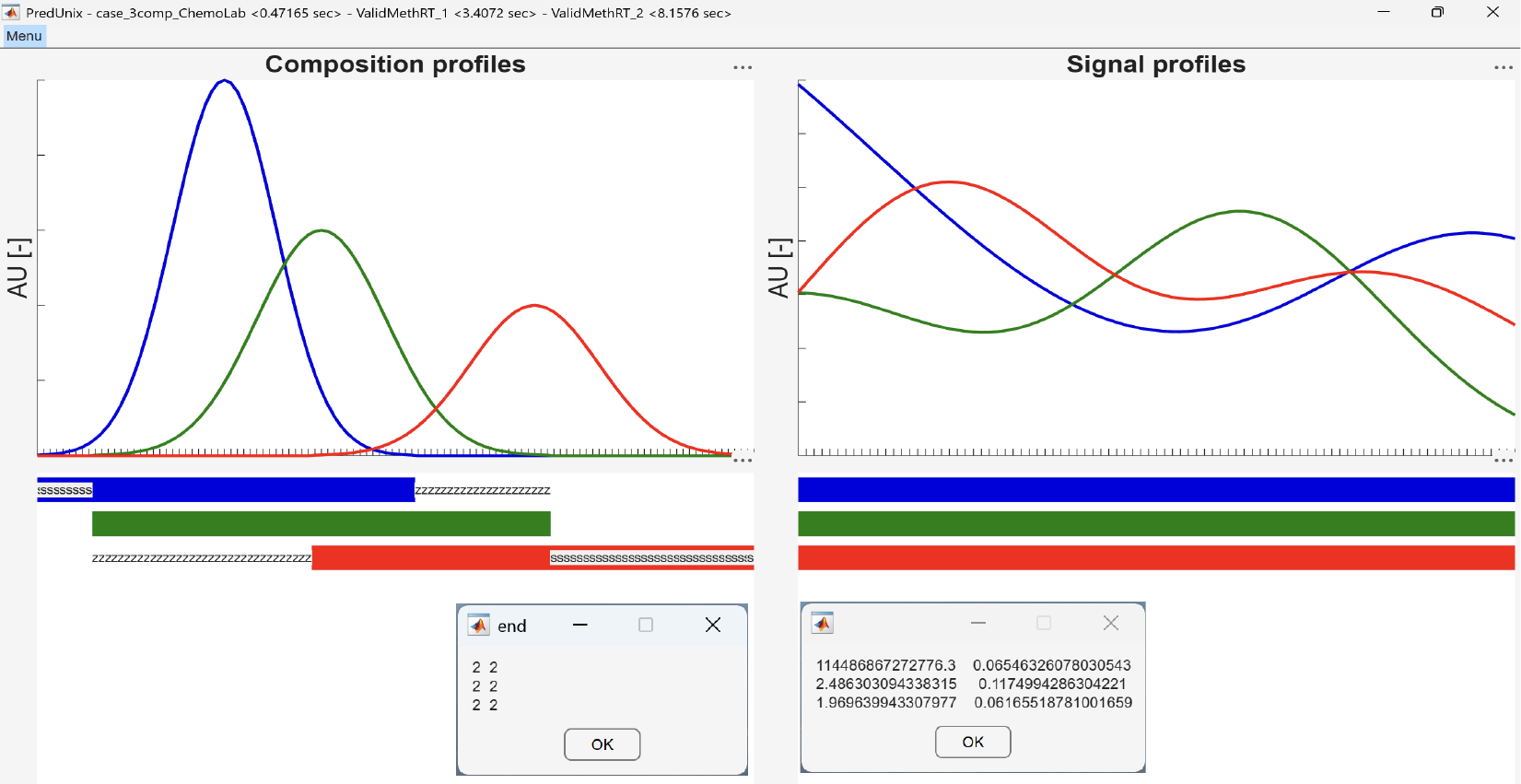}
	\caption{
    Constructive counterexample illustrating why satisfaction of Manne’s Theorems 1 and 2 alone does not establish unique bilinear resolution. For the selected blue/red analyte, the concentration/composition profiles contain the apparent zero-component region required by Theorem 1 and the selective subwindow required by Theorem 2. Nevertheless, the corresponding spectral/signal profiles contain neither a zero-component region nor a selective subwindow that fixes the relevant feasible boundary. Consequently, the concentration-side conditions do not reduce the feasible set to a single factorization. The green component contains isolated zero values, but these do not form a zero-component subwindow and therefore do not satisfy the relevant condition. The inset validation results illustrate that the constructed data set is non-unique: much slower two methods (for a unique profile, the corresponding value should be zero in the inserted tables) show that the PredUnix algorithm worked properly and fast. 
    }\label{fig_ManneThCountExample}
\end{figure*}

Figure 8 of their paper demonstrated a counterexample against Manne's theorems, and in mathematics only one counterexample is enough to disprove any theorems. However, we can show another counterexample on Fig.~\ref{fig_ManneThCountExample} to show how one can create synthetic data in general which fulfills the conditions of the Manne's theorems, but there is not unique solution. The details on the mentioned PredUnix and its validation algorithms will be published elsewhere.

\cite{RAJKO2015} also defined the uniqueness levels as non-uniqueness, fractional uniqueness, partial uniqueness and full uniqueness, and they showed this concept through several depicted examples.

An unfortunate interpretation was published by~\cite{Akbari2013}, and some part of the conclusion is repeated here: {\it''In general, the successful resolution depends on the structure of the dataset and is related to the presence of selective regions, where only one component exists or one component is absent from a mixture. Appropriate imposing of the zero regions constraint can guarantee the unique calculation of profiles in iterative soft-modeling methods. The presence of selective region in one mode of the data set may not be sufficient for obtaining
the unique solution without imposing the zero regions constraint.''} In our view, the paper correctly illustrates that the use of local-rank and selectivity information alone does not generally establish uniqueness. In the investigated three-component example, the feasible-solution analysis displays three non-degenerate feasible regions, one for each component. Therefore, under the stated nonnegativity conditions, the decomposition is not unique. Importantly, if a profile in that example is regarded as fulfilling the conditions of Manne’s theorems, the example itself constitutes an additional counterexample to the interpretation that these conditions are sufficient for uniqueness in an iterative bilinear decomposition. The central issue is that local-rank and selectivity conditions may identify a profile that can be calculated by a particular direct procedure, but they do not by themselves ensure that the corresponding profile is the sole feasible profile within the complete set of nonnegative bilinear decompositions.

\section{General rule for uniqueness (GRU)}\label{GRU}

\cite{gru2020} proposed the general rule for uniqueness to unify all the different information that led to a unique solution in one framework: {\it ''The duality concept is a powerful tool in explaining and interpreting numerous features of MCR methods. Based on this
concept, a general rule was employed to explain the unique solutions of MCR methods as a result of applying different constraints. The GRU states that any information, able to fix the subspace of all complementary components in one space, results in a unique solution for the analyte in the other space, which is a powerful use of the duality concept. This generalization is presented for constraints such as trilinearity, equality, zero concentration region, correspondence, local rank, non-negativity under DBU conditions, and even to extract the NAS. GRU is a proper index to find the conditions and information in different chemical systems probably leading to a unique solution in applying MCR methods.''}

\section{Duality}\label{Duality}
\cite{HENRY2005} introduced the duality relationship, but described it only for multivariate receptor modeling of compositional data of airborne pollution. \cite{RAJKO2006} generalized Henry concept, giving detailed matrix-based and convex geometrical proof to introduce the concept of the minimal constrained (using only nonnegativity) duality.


\begin{align}\label{eq_ex_align}
	\text{V-space:\ } 
	\underbrace{\underset{I \times N}{\mathbf{X}}}_{\text{points}}  \text{and\ }
	\underset{J \times N}{\mathbf{Y}}
	\underset{N \times N}{\mathbf{D}^{-1}}
	\underset{N \times 1}{\mathbf{z}}=
	\underbrace{\underset{J \times N}{\mathbf{V}}
	\underset{N \times 1}{\mathbf{z}} \geq 
	\underset{J \times 1}{\mathbf{0}}}_{
	\text{half subspaces}}
	\nonumber \nopagebreak\\
	\rotatebox{-12}{$\xleftrightarrow{\hspace{43mm}}$}\hspace{9mm} \nonumber\\[-2.7\baselineskip]
	\rotatebox{12}{$\xleftrightarrow{\hspace{43mm}}$}\hspace{9mm}
	\nonumber \nopagebreak\\
	\text{U-space:\ } 
	\underbrace{\underset{J \times N}{\mathbf{Y}}}_{
	\text{points}} \text{and\ }
	\underset{I \times N}{\mathbf{X}}
	\underset{N \times N}{\mathbf{D}^{-1}}
	\underset{N \times 1}{\mathbf{z}}=
	\underbrace{\underset{I \times N}{\mathbf{U}}
	\underset{N \times 1}{\mathbf{z}} \geq 
	\underset{I \times 1}{\mathbf{0}}}_{
	\text{half subspaces}} \nopagebreak
\end{align}
The duality between points and half subspaces (considering only the equalities, hyperplanes), denoted by the arrows, can be read from Eq.~\eqref{eq_ex_align}. As \cite{RAJKO2006} stated {\it ''Thus, we can see that the points and the hyperplanes are
duals of each other. Moreover, as many points are defined in one subspace (say V-space), the same number of hyperplanes will be defined in the other one (say U-space).''}

\cite{RAJKO2006} gave a short proof as well: {\it ''The pyramids are called polyhedral cones by mathematicians (\cite{stoer2012convexity}) and every polyhedral cone is the solution set of a homogeneous system of inequalities, like $\underset{I \times N}{\mathbf{U}}
\underset{N \times 1}{\mathbf{z}} \geq 
\underset{I \times 1}{\mathbf{0}}$. For
an arbitrary polyhedral cone $S \subseteq \mathbb{R}^N$, its polar cone can be defined as
\begin{equation}\label{eq_polar}
	S^p \stackrel{\text{def}}{=} \Bigl\{ \mathbf{y} \in \mathbb{R}^N | \langle \mathbf{y},\mathbf{z} \rangle \ge 0, \textnormal{\ \ \ for all\ } \mathbf{z} \in S\Bigr\},
\end{equation}
where $\langle . \rangle$ means the inner product of two vectors, here defined by $\langle \mathbf{y},\mathbf{z} \rangle = \underset{1 \times N}{\mathbf{y}^\textnormal{T}}
\underset{N \times N}{\mathbf{D}^{-1}}
\underset{N \times 1}{\mathbf{z}}$, $\mathbf{D}$ is a diagonal matrix (practically the singular values of an appropriate matrix are in the diagonal).
A face of $S$ is a cone $F \subset S$ such that for all $\mathbf{z} \in F$, if
$\mathbf{z} = \mathbf{z}_1+\mathbf{z}_2$ with $ \mathbf{z}_1,\mathbf{z}_2 \in S$ then $ \mathbf{z}_1,\mathbf{z}_2 \in F$. A face of
dimension 1 is called an extremal ray of $S$, while a face of codimension 1 is called a facet of $S$. The following statement is used for making complete the proof: the dual of a facet of $S$ is an extremal ray of $S^p$, which follows from the definition Eq.~\eqref{eq_polar}.
Thus, the inner polyhedral cone in V-space is the polar cone of outer one in U-space and vice versa; and the outer polyhedral cone in V-space is the polar cone of inner one in U-space and vice versa. Because of the special property of the cones and its polar cones, it is proved that the points generating the convex hull of the inner pyramid in one space are the dual boundary hyperplanes of the outer pyramid in the other space.''}

\begin{figure*}
	\centering
	\includegraphics[width=1\textwidth]{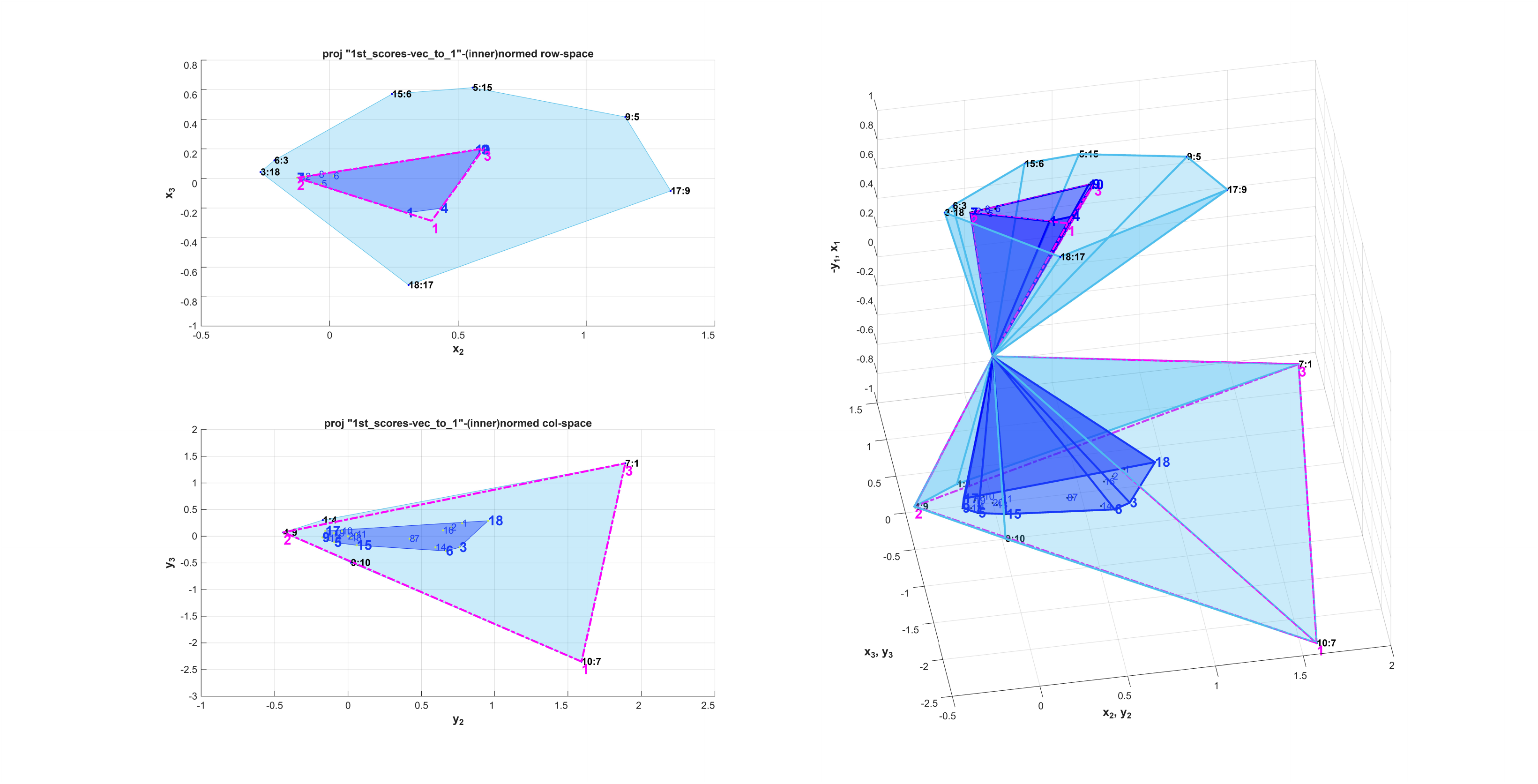}
	\caption{
    Visual representation of the minimal constrained duality for a three-component system. 
    Left: normalized concentration- and spectral-mode subspaces, showing the inner and outer polygons. The outer polygons delimit feasible normalized profiles under nonnegativity. Right: the equivalent polyhedral-cone representation, in which boundary hyperplanes in one mode correspond through polarity to extremal rays in the dual mode.
    The vertices of the pink triangles represent the pure profiles. \href{https://drive.google.com/file/d/1JBpCBrXi4YDazQzw6ltiSaNwl8qJcT-S/view?usp=drive_link}{Rotatable MATLAB figure file} is provided as Supplementary Material.
    }\label{fig_DualityPyramid}
    
\end{figure*}

Fig.~\ref{fig_DualityPyramid} depicts the minimal constrained duality for a three-component chemical system published by \cite{HENRY1990}. We can interpret the two normalized subspaces for the concentration and spectral ways (left side pane), and the polyhedral cones and its duals (right side pane). The known profiles are also indicated with pink dashed-dotted lines. 

\begin{figure*}
	\centering
	\includegraphics[width=1\textwidth]{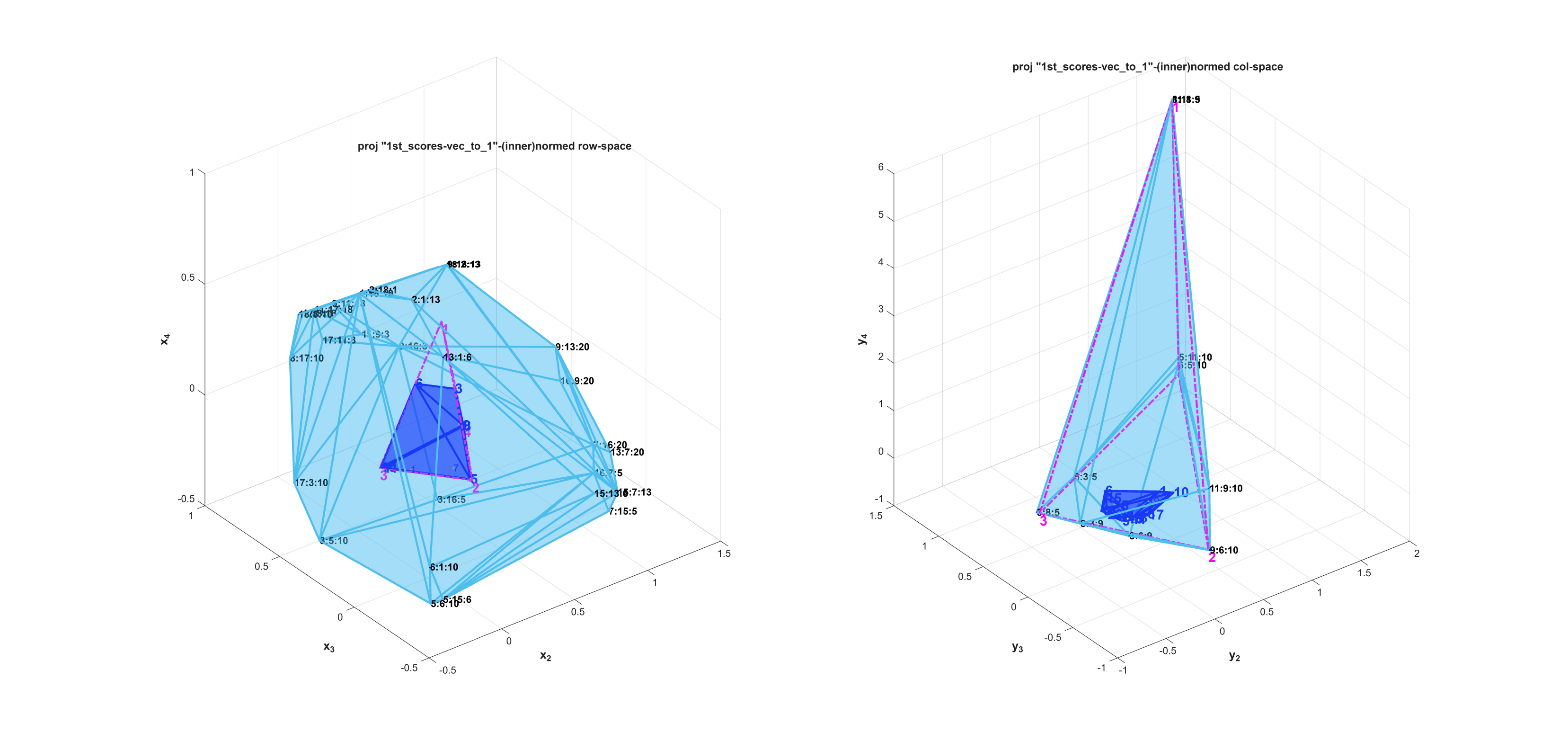}
	\caption{Visual representation of the minimal constrained duality for a four-component system showing the inner and outer polyhedra.
    Left: normalized concentration-mode subspace. Right: normalized spectral-mode subspace. 
    The vertices of the pink tetrahedra represent the pure profiles. \href{https://drive.google.com/file/d/1FIl7hHGgZ6hfP7My8K2e4HLYUTHirbhl/view?usp=drive_link}{Rotatable MATLAB figure file} is provided as Supplementary Material.
    }\label{fig_Duality4comp}
\end{figure*}

From four components, the representation of the polyhedral cone--polar cone systems is quite difficult, thus we show only the two normalized 3D subspaces in Fig.~\ref{fig_Duality4comp} as illustrations using a 4-component chemical system provided by \cite{KIM1999}.

\cite{ABDOLLAHI2011} published a review-like report on the uniqueness. First of all, they used the term 'rotation' ambiguity, instead of the previously coined and widely accepted 'rotational' one. Rotation is for if the object can rotate, rotational is for if somebody makes the object rotate. They also stated {\it ''The possible presence of rotation ambiguities in the analysis of experimental data sets is implicit to all MCR methods, and it is an intrinsic limitation of them.''} However, the assumed bilinear model and the possible theoretical decomposition(s) have these ambiguities and not the methods. The MCR methods are for only estimating the parameters of the bilinear model. The decomposition with any invertible transformation matrix $\mathbf{T}$ and its inverse can yield a new feasible solution with the same optimal properties:
\begin{equation}\label{eq_svdTrans}
	\underset{I \times J}{\mathbf{R}} = \underset{I \times N}{\mathbf{U}} \underset{N \times N}{\mathbf{D}}
	\underset{N \times N}{\mathbf{T}}\ \ \ 
	\underset{N \times N}{\mathbf{T}^{\textnormal{-1}}}
	\underset{N \times J}{\mathbf{V}^{\textnormal{T}}} = 
	\underset{I \times N}{\widetilde{\mathbf{U}}} \underset{N \times N}{\widetilde{\mathbf{D}}}\ \ \ 
	\underset{N \times J}{\widetilde{\mathbf{V}}^{\textnormal{T}}}
\end{equation}

 {\it ''There are three types of ambiguities in MCR methods, permutation, intensity and rotation ambiguities''} -- however, the three ambiguities (label switching, scaling and skew transformation) together can be treated by only one invertible transformation matrix $\mathbf{T}$:
 \begin{equation}\label{eq_T3types}
 	\underset{N \times N}{\mathbf{T}} = {\mathbf{T}_\textnormal{skew tr.}} 
 	{\mathbf{T}_\textnormal{perm.}} {\mathbf{T}_\textnormal{int.}} = \underset{N \times N}{\mathbf{T}_\textnormal{s}} 
 	\underset{N \times N}{\mathbf{T}_\textnormal{p}} \underset{N \times N}{\mathbf{T}_\textnormal{i}},
 \end{equation}
 and the following holds:
 \begin{equation}\label{eq_Ttogather}
 	\underset{I \times J}{\mathbf{R}} = 
 	\underset{I \times N}{\mathbf{C}} \underset{N \times J}{\mathbf{S}^\textnormal{T}} =
 	\underbrace{\underset{I \times N}{\mathbf{C}}
 	{\mathbf{T}_\textnormal{s}} 
 	{\mathbf{T}_\textnormal{p}} {\mathbf{T}_\textnormal{i}}}_{\underset{I \times N}{\widetilde{\mathbf{U}}} \underset{N \times N}{\widetilde{\mathbf{D}}}}\ \ \ 
 	\underbrace{{\mathbf{T}_\textnormal{i}^\textnormal{-1}}
 	{\mathbf{T}_\textnormal{p}^\textnormal{T}}
 	{\mathbf{T}_\textnormal{s}^\textnormal{-1}}
 	\underset{N \times J}{\mathbf{S}^\textnormal{T}}}_{\underset{N \times J}{\widetilde{\mathbf{V}}^{\textnormal{T}}}}.
 \end{equation}
 
 {\it ''Although the nonuniqueness problem is ubiquitous to all MCR methods, it can be alleviated and totally avoided in some cases by means of an intelligent use of the data structure and of appropriate constraints.''} -- Again, the rotational ambiguities belong to the bilinear model (including the measurement data) and not to the methods! The methods just estimate the parameters and provide only one solution after several iteration steps, and this particular/single solution is not generally unique. The constraints are useful if they are based on chemical/physical facts (nonnegativity, unimodality, kinetic hard models etc.), and not just mathematical ones, e.g., independency, maximum sparseness, maximum determinant/volume criterion etc.
 
 {\it ''The final goal of this paper is therefore, to contribute to the understanding of MCR methods and to eliminate the misconceptions and misunderstandings frequently appearing in the literature about better performances of one particular MCR method over another [29,30].''} -- 
 Ref. 30 (R. Rajko Anal, Chem. 82 (2010) 8750–8752.) does not contain any misconceptions and it is not about better performance of a particular MCR method. Minimum Volume Simplex Analysis (MVSA) and related methods used by Lopes et al. (Ref. 29 of \cite{ABDOLLAHI2011}) were developed for compositions when the following should be fulfilled: “A necessary condition for this idea to work is that each facet of the simplex contains at least $p - 1$ spectral vectors.” And as Lopes et al. wrote in their reply to the comment (Ref. 30 of \cite{ABDOLLAHI2011}) “for each of the $p$ pure materials (endmembers), there should exist in the data at least $p - 1$ pixels where the abundance of this pure material is zero” and this means that the tablets are not homogeneous at all, which is an elementary mandatory for official pharmaceutical productions. 
 Thus it is evident that for both of the data presented in the paper, MVSA will 
 be unusable. Moreover it was illustrated using previously published data that even the simplexes in the concentration space can be outside the feasible regions obtained by MVSA and related methods. Moreover, MVSA seems to be strictly mathematically constrained method, which should be avoided, as we discussed it above. 
 
 Why MVSA should be considered for investigating MCR and uniqueness? 
 In the NMF and hyperspectral-unmixing literature, e.g., \cite{Lin2015,Fu2018,Than2024,barbarino2025,nguyen2025,qi2025,Qi2026,Walinjkar2026}, identifiability usually denotes uniqueness of the complete factorization, up to permutation and scaling, and a minimum-volume criterion is introduced to select an identifiable factorization under additional assumptions. In MCR, however, the information content of a decomposition is not exhausted by this all-or-nothing classification. A chemical component may be uniquely resolved in one mode, or in both modes, although other components remain non-unique. We therefore distinguish fractional uniqueness, partial uniqueness, and full uniqueness, see~\cite{RAJKO2015}. Minimum-volume selection may produce a particular factorization, but it should not be conflated with a profile-wise proof of uniqueness under minimal nonnegativity constraints.
 This chemometric distinction has a direct counterpart in exact NMF: partial identifiability (see the details later in Section~\ref{PartUniq}) results establish conditions under which individual factors, or successively more factors, are uniquely identifiable even when global identifiability of the entire factorization is not established.
 
 {\it ''The method based on the duality concept [37,38] was finally preferred in this work for its simplicity and speed, allowing the direct analytical calculation of the limits or boundaries of the band of feasible solutions.''}
  -- There is nothing about the exact equations derived by using the duality concept.
 
 {\it ''However, explicit analytical calculation of the set of feasible solutions and of their possible boundaries for a three component system is more difficult and involved than it is for a two component system.''} -- At that time, two such kinds of algorithms were published: one based on computational geometrical tools, and an other based on duality concept. The version based on the duality concept is more sensitive to the noise; however in the paper noiseless cases were shown, and in these cases either algorithms could have been used.
 
 {\it ''After appropriate normalization (see Eq. (7)), these values can be projected to a hyperplane (a r-1 dimensional subspace)...''} -- Other types of the appropriate normalization are possible, and Rajko’s Borgen norms paper (\cite{RAJKO2009}) should have been treated and cited here.
 
 {\it ''For a three component system, the method of calculation of the set of feasible solutions based on the grid search method was preferred because it is easier to implement and understand than the methods based on the duality concept or on Borgen plots.''} -- At that time, the feasible solutions of Eq.~\eqref{eq_Ttogather} could be calculated and represented by Borgen plot method alone. 
 The grid search method, the simplex based searching method [Ref. 31-33] etc. are just approximations for the Borgen plot, but not alternatives of it! 
 They only try to get approximate information and try to reproduce the information of the theoretical Borgen plot which contains the theoretically proper information alone.
 With an approximation method, it is very harmful to draw considerable conclusions. We can guess several useful things, we can get usable hints, but the proofs should have been done by theoretical methods.

\cite{Sawall2012} introduced complementarity and coupling theorems utilizing the partial knowledge of the factors. Surprisingly, they stated {\it ''It is worth to note that the complementarity-coupling theorems are different from the duality results as introduced by Henry [13] for multivariate receptor modeling and discussed by Rajkó [24] in the SMCR context. This duality approach uses the non-negativity constraints to find restrictions on the feasible regions and works with the "external" matrices $\tilde{U}$ and $\tilde{V}$ in (4). In contrast to this, the complementarity-coupling approach is based on the partial knowledge of the factors $C$ or $A$ and uses the coupling through the 
"inner" matrix pair $T$ and $T^{-1}$ in (4) in order to derive restrictions on the feasible regions.''} The complementarity–coupling formulation and the minimal constrained duality formulation emphasize different mathematical representations of related geometric restrictions. Under nonnegativity constraints in both modes, the correspondence between faces of a polyhedral cone and faces of its polar cone provides a direct geometric framework for relating these restrictions.

50 years ago, \cite{Tam_1976} proved that not only the above mentioned extremal ray--facet connection exists between the polyhedral cone and its polar, but it can be generalized between any face with dimension $n$ (polyhedral cone) and face with codimension $n$, i.e. dimension $N-n$ (polar cone), cf. Eq.~\eqref{eq_polar}, and the explanations around it. In the polyhedral cone, we can fix a face with dimension $n$, if we know the corresponding profiles in advance, thus the face with codimension $n$ in the polar cone will be fixed as well. \cite{HENRY2005} stated that {\it ''It
is a mathematical theorem that every finite dimensional
vector space is isomorphic to its dual space [4]. It is also a mathematical fact that there is no "natural" isomorphic mapping between the vector space and its dual in the sense that the mapping arises in a natural way and is not dependent on the specific set of basis vectors used to define the space [4]. This paper identifies a specific isomorphism between a vector space important in receptor modeling and its dual space. Since there is no natural choice, this isomorphism is somewhat arbitrary and it is possible to define other duality relationships that would work just as well.''} Because \cite{Sawall2012} used general vector space related to such matrices which can contain negative elements as well, their used isomorphism will be arbitrary, and there is no sophisticated reasoning why we have to choose that special isomorphism. If  nonnegativity constraint for both ways of a data matrix is applied, the complementarity-coupling approach will be totally equivalent to the faces of the polyhedral cone and its dual in convex geometry world, i.e., to the minimal constrained duality, as it was introduced by \cite{RAJKO2006}. In addition, the coupling theorem is related to some kind of normalization, however one could have used other type of normalization, e.g., the above mentioned Borgen norms introduced by \cite{RAJKO2009}, which are the generalization of the $L_1$ norm. 

We think, that the convex geometry picture is much clearer and understandable for an average chemometrician than using mathematical rigorous linear algebra formulas, theorems and proofs. 


\section{Particular solution based uniqueness via chemometrical and via mathematical way}\label{PartUniq}
\cite{AKBARILAKEH2022} published the paper on predicting the uniqueness based on only particular non-negative profiles estimated by multivariate curve resolution methods which is an elegant simplification and generalization of the data-based uniqueness concept:   

{\it ''The uniquely recovered profiles can be diagnosed easily and without any calculations from the results of the bilinear decomposition of chemical data using the MCR-ALS (or analogous) method under non-negativity constraints. These profiles are the real solutions of the system that are obtained and detected with minimal information. The strategy for detecting all the uniquely resolved profiles of an MCR-ALS (or analogous) analysis, by applying only non-negativity constraints, can be summarized in the following steps (our proposed procedure):

1. Check for all profiles the fractional uniqueness based on DBU.

2. Discard species that are unique in both modes, i.e., the partial unique components, and recheck the remaining system for additional fractional uniqueness based on DBU.

3. Check the number of uniquely resolved components for both modes. If this number is equal to $\#comp-1$ (one less than the number of components) in one mode, then based on GRU, the complement component in the other mode is also uniquely resolved.''}

This is the proper reformulation of the profile-based uniqueness (restricted DBU), what was tried to work out by Manne. Even some useful thumb-rules were given about identifying the non-unique resolutions: 

{\it ''Based on the above-mentioned strategy, several shortcuts are extracted for rapidly predicting some of the non-unique solutions of an MCR-ALS (or analogous) analysis.
	
a) When none of the resolved profiles contain any zero-regions, there is no unique solution.

b) When none of the resolved profiles of a specific mode (e.g., concentration mode) contain any zero-regions, there is no unique solution.

c) When none of the profiles of a species contain any zero-regions, there is no unique solution.

d) When there are no zero-regions in a specific profile, that profile is not unique.
	
As mentioned before, the zero-regions in concentration or spectral modes are defined after considering a noise level for each system. The above statements do not indicate all possible situations for a non-unique solution, but give quick instructions for finding some cases of non-uniqueness.''}

Later, related to these chemometrical results, \cite{Gillis2023} provided the rigorous mathematical proves for several connected theorems:

{\it ''In this paper, we have provided the following partial identifiability results for exact NMF:
	
\textbullet\ a rigorous description and proof of the restricted DBU theorem (Theorem 6);

\textbullet\  a new partial identifiability result based on the geometric interpretation of
	the restricted DBU theorem (Theorem 7);
	
\textbullet\  a sequential approach to guarantee the identifiability of more factors (Theorem
	8).''}



\section{Conclusion}\label{Conclusion}
The same distinction between feasible resolution and demonstrable uniqueness is relevant beyond bilinear MCR. In nonnegative three-way decomposition, trilinearity can substantially reduce the feasible set; however, it does not replace an analysis of the data structure. In our earlier results for two- and three-component three-way arrays, \cite{RAJKO2017} showed that a unique constrained decomposition can occur even when Kruskal’s rank inequality is not fulfilled, provided that the relevant selective regions and zero-pattern information are present in the profiles. Conversely, rank-based conditions alone do not exhaust the information controlling identifiability under nonnegativity constraints.
This observation supports the central message of the present work: uniqueness should not be inferred merely because an algorithm returns a chemically plausible factorization, nor solely because a local-rank or selectivity argument permits calculation of a particular profile. The relevant question is whether the available information collapses the feasible region to one point. In bilinear MCR, DBU expresses this requirement geometrically through the coincidence of the appropriate inner and outer feasible-set boundaries. GRU provides the complementary interpretation: information that fixes the subspace of all complementary components in one mode yields a unique profile in the other mode.
For practical MCR work, the resulting procedure is straightforward. First, obtain a physically credible nonnegative factorization and distinguish it from a proof of uniqueness. Second, inspect the data structure for chemically justified zero-component regions, selectivity, and other information that restricts the feasible set. Third, assess profile-oriented restricted DBU and, where appropriate, use the general rule for uniqueness to identify complementary uniquely resolved profiles. This workflow allows the analyst to distinguish non-unique, fractionally unique, partially unique, and fully unique results, rather than treating every converged MCR solution as uniquely resolved.
Thus, the purpose of this re-tuned account is not to replace mathematical uniqueness theory, but to align its interpretation with the practical language of chemometrics. Manne’s resolution theorems, data-based uniqueness, general uniqueness rules, minimal constrained duality, and particular-solution diagnostics all address aspects of the same question: what information in a given chemical data set is sufficient to eliminate the remaining admissible alternatives? A reliable answer requires examining the complete feasible geometry of the decomposition rather than relying on an isolated rank condition, a selective region, or an algorithmic solution.

And in the end, according to the principle of explosion or {\it ex falso quodlibet}, if an axiomatic system contains even a single contradiction (e.g., accepting a false theorem), any statements (including wrong ones) can be logically derived from it, and it will be possible to ruin some parts or even the whole system.

\printcredits

\bibliographystyle{cas-model2-names}

\bibliography{cas-refs}



\end{document}